\documentstyle[12pt]{article}

\expandafter\ifx\csname mathrm\endcsname\relax\def\mathrm#1{{\rm #1}}\fi

\makeatletter
\if@twoside \oddsidemargin -17pt \evensidemargin 00pt
\else \oddsidemargin 00pt \evensidemargin 00pt
\fi
\expandafter\ifx\csname mathrm\endcsname\relax\def\mathrm#1{{\rm #1}}\fi

\makeatother

\def\beq{\begin{equation}}
\def\eeq{\end{equation}}
\def\beqar{\begin{eqnarray}}
\def\eeqar{\end{eqnarray}}
\def\barr#1{\begin{array}{#1}}
\def\earr{\end{array}}
\def\bfi{\begin{figure}}
\def\efi{\end{figure}}
\def\btab{\begin{table}}
\def\etab{\end{table}}
\def\bce{\begin{center}}
\def\ece{\end{center}}
\def\nn{\nonumber}
\def\disp{\displaystyle}
\def\text{\textstyle}

\def\ga{\gamma}
\def\de{\delta}

\def\si{\sigma}
\def\ie{i\varepsilon}

\def\reffi#1{\mbox{Fig.~\ref{#1}}}

\def\refta#1{\mbox{Tab.~\ref{#1}}}
\def\refta#1{\mbox{Table~\ref{#1}}}

\def\refse#1{\mbox{Sect.~\ref{#1}}}

\def\refapp#1{\mbox{App.~\ref{#1}}}

\newcommand{\GeV}{\unskip\,\mathrm{GeV}}

\newcommand{\TeV}{\unskip\,\mathrm{TeV}}

\newcommand{\pba}{\unskip\,\mathrm{pb}}
\renewcommand{\O}{{\cal O}}

\def\mathswitchr#1{\relax\ifmmode{\mathrm{#1}}\else$\mathrm{#1}$\fi}

\newcommand{\PW}{\mathswitchr W}
\newcommand{\PZ}{\mathswitchr Z}

\newcommand{\Pe}{\mathswitchr e}
\newcommand{\Pne}{\mathswitch \nu_{\mathrm{e}}}

\newcommand{\Pep}{\mathswitchr {e^+}}
\newcommand{\Pem}{\mathswitchr {e^-}}

\newcommand{\PWm}{\mathswitchr {W^-}}

\def\mathswitch#1{\relax\ifmmode#1\else$#1$\fi}

\newcommand{\MW}{\mathswitch {M_\PW}}

\newcommand{\MZ}{\mathswitch {M_\PZ}}

\newcommand{\Me}{\mathswitch {m_\Pe}}

\newcommand{\sw}{\mathswitch {s_\PW}}
\newcommand{\cw}{\mathswitch {c_\PW}}

\newcommand{\M}{{\cal {M}}}

\def\Li{\mathop{\mathrm{Li}_2}\nolimits}

\def\solid{\raise.9mm\hbox{\protect\rule{1.1cm}{.2mm}}}

\def\draftdate{\relax}
\def\mda{\relax}
\def\mua{\relax}
\def\mla{\relax}
\def\mpar#1{\relax}
\def\draft{
\def\draftdate{\today}
\def\mpar##1{\marginpar{\hbadness10000\sloppy\boldmath\bf##1}%
                      \typeout{marginpar: \noexpand##1}\ignorespaces}
\def\mda{\mpar{\hfil$\downarrow$\hfil}}
\def\mua{\mpar{\hfil$\uparrow$\hfil}}
\def\mla{\marginpar[\boldmath\hfil$\rightarrow$\hfil]%
                   {\boldmath\hfil$\leftarrow $\hfil}%
                    \typeout{marginpar: $\leftrightarrow$}\ignorespaces}
}

\makeatletter
\let\@eqnsel = \hfil

\def\eqnarray{\stepcounter{equation}\let\@currentlabel=\theequation
\global\@eqnswtrue
\global\@eqcnt\z@\tabskip\@centering\let\\=\@eqncr
$$\halign to \displaywidth\bgroup\hskip\@centering
  $\displaystyle\tabskip\z@{##}$\@eqnsel&\global\@eqcnt\@ne
  \hskip 2\arraycolsep \hfil${##}$\hfil
  &\global\@eqcnt\tw@ \hskip 2\arraycolsep $\displaystyle\tabskip\z@{##}$\hfil
   \tabskip\@centering&\llap{##}\tabskip\z@\cr}
\def\appendix{\par
 \setcounter{section}{0} \setcounter{subsection}{0}
 \def\thesection{\Alph{section}}}

\makeatother

\newcommand{\egnwezeg}{$\Pem\ga\to\PWm\Pne$, $\Pem\PZ$, $\Pem\ga$}
\newcommand{\egnwez}{$\Pem\ga\to\PWm\Pne$, $\Pem\PZ$}

\newcommand{\egeg}{\mathswitch{\Pem\ga\to\Pem\ga}}

\newcommand{\egnwgezg}{$\Pem\ga\to\PWm\Pne\ga$, $\Pem\PZ\ga$}

\newcommand{\egezg}{$\Pem\ga\to\Pem\PZ\ga$}
\newcommand{\egegg}{$\Pem\ga\to\Pem\ga\ga$}

\def\born{{\mathrm{Born}}}

\def\QED{{\mathrm{QED}}}
\def\NC{{\mathrm{NC}}}

\newcommand{\lsim}
{\;\raisebox{-.3em}{$\stackrel{\displaystyle <}{\sim}$}\;}

\begin{document}
\thispagestyle{empty}
\def\thefootnote{\fnsymbol{footnote}}
\setcounter{footnote}{1}
\null
\hfill BI-TP 93/61 
\vskip 1cm
\vfil
\begin{center}
{\Large \bf Full $\O(\alpha)$ Radiative Corrections \\
to High-Energy Compton Scattering%
\footnote{Work partially supported by the Bundesminister f\"ur Forschung 
und Technologie, Bonn, Germany.}
\par} \vskip 5em
{\large
{\sc S.\ Dittmaier}  \\[1ex]
{\it Fakult\"at f\"ur Physik, Universit\"at Bielefeld, Germany}
\par} \vskip 1em
\end{center} \par
\vskip 4cm
\vfil
{\bf Abstract} \par
Using computer-algebraic methods we derive compact analytical expressions
for the virtual electroweak radiative corrections to polarized Compton
scattering. Moreover the helicity amplitudes for double Compton
scattering, which prove to be extremely simple in terms of Weyl-van der
Waerden spinor products, are presented for massless electrons. The
inclusion of a finite electron mass is described, too. Finally numerical
results both for the purely photonic and the full $\O(\alpha)$
electroweak corrections, which turn out to be of the order of $5-10\%$, 
are discussed for energies ranging from 10\GeV\ to 2\TeV.
\par
\vskip 2.3cm
\noindent November 1993 \par
\null
\setcounter{page}{0}
\clearpage
\def\thefootnote{\arabic{footnote}}
\setcounter{footnote}{0}

\section{Introduction}

Since the experimental discovery of the reaction \egeg\ by Compton
\cite{co23} in 1923 this so-called Compton scattering has been of
continuous theoretical interest. Among the numerous work on this
subject we just mention the most basic representatives, e.g.\ the
lowest-order cross-section calculated by Klein and Nishina
\cite{klni29}, the virtual and real soft-photonic QED radiative 
corrections (RCs) by Brown and Feynman \cite{brfe52} as well as the 
hard-photonic corrections -- also called double Compton scattering 
-- by Mandl and Skyrme \cite{mask52}. Recently these results have been
completed by the virtual electroweak RCs \cite{egeg} within the
Glashow-Salam-Weinberg model as part of a more general treatment of
gauge-boson production in electron-photon collisions 
\cite{sddis}.

The relatively clean environment of electron-photon collisions will
provide an opportunity for further precision tests of the electroweak
standard model, which are complementary to the ones obtained from
electron-positron scattering, in the future. As proposed 
in \cite{plc} high-energetic photon beams can be produced via Compton 
backscattering of laser light off high-energy electrons. Moreover
elastic electron-photon backscattering is well-suited as luminosity
monitor for such an electron-photon collider in analogy to Bhabha
forward scattering in electron-positron colliders. Consequently
the Compton process represents one of the most important processes
in this context.

In \cite{egeg} the analytical results for \egeg\ are related 
to the generic ones obtained for \egnwez\ \cite{egnwez} rendering the
analytical expressions quite untransparent. On the other hand the
kinematical simplicity of Compton scattering for energies at the
electroweak scale ($\MW\gg\Me$) promises a comparably simple structure
for the electroweak RCs although intermediate steps of the calculations
are lengthy. Therefore we have applied the computer-algebra packages
{\it FeynArts} \cite{fa} and {\it FeynCalc} \cite{fc} for generating the 
one-loop amplitudes for \egeg\ and reducing them to scalar integrals,
respectively. The scalar one-loop integrals have been evaluated by
standard methods \cite{scalar}. The rather transparent and short results
obtained this way form the first part of this work.

In a second step we deal with double Compton scattering. Following
closely the procedure of \cite{egnwgezg}, where the hard-photonic
bremsstrahlung to \egnwez\ is discussed, we use the Weyl-van der Waerden
spinor formalism for the construction of very compact helicity
amplitudes in the case of non-collinear photon emission. The finite-mass 
effects of the electron are included afterwards.

Finally we present numerical results for the pure (virtual and real)
photonic as well as for the full $\O(\alpha)$ RCs to the integrated Compton
cross-section both for polarized and unpolarized particles for energies
ranging from 10\GeV\ to 2\TeV.

The paper is organized as follows: In \refse{notcon} we set some
conventions and give the polarized Born cross-sections. The 
virtual electroweak RCs and the real soft-photonic corrections are
presented in \refse{virc}. Section \ref{hbrc} completes these analytical
results by the hard-photonic bremsstrahlung. We conclude with a 
discussion of the numerical evaluations in \refse{results}. Finally the
appendix provides completing analytical expressions.

\section{Notation and lowest-order cross-section}
\label{notcon}

\begin{figure}
\begin{center}
\begin{picture}(10.5,2)
\put(-2.5,-15){\includegraphics{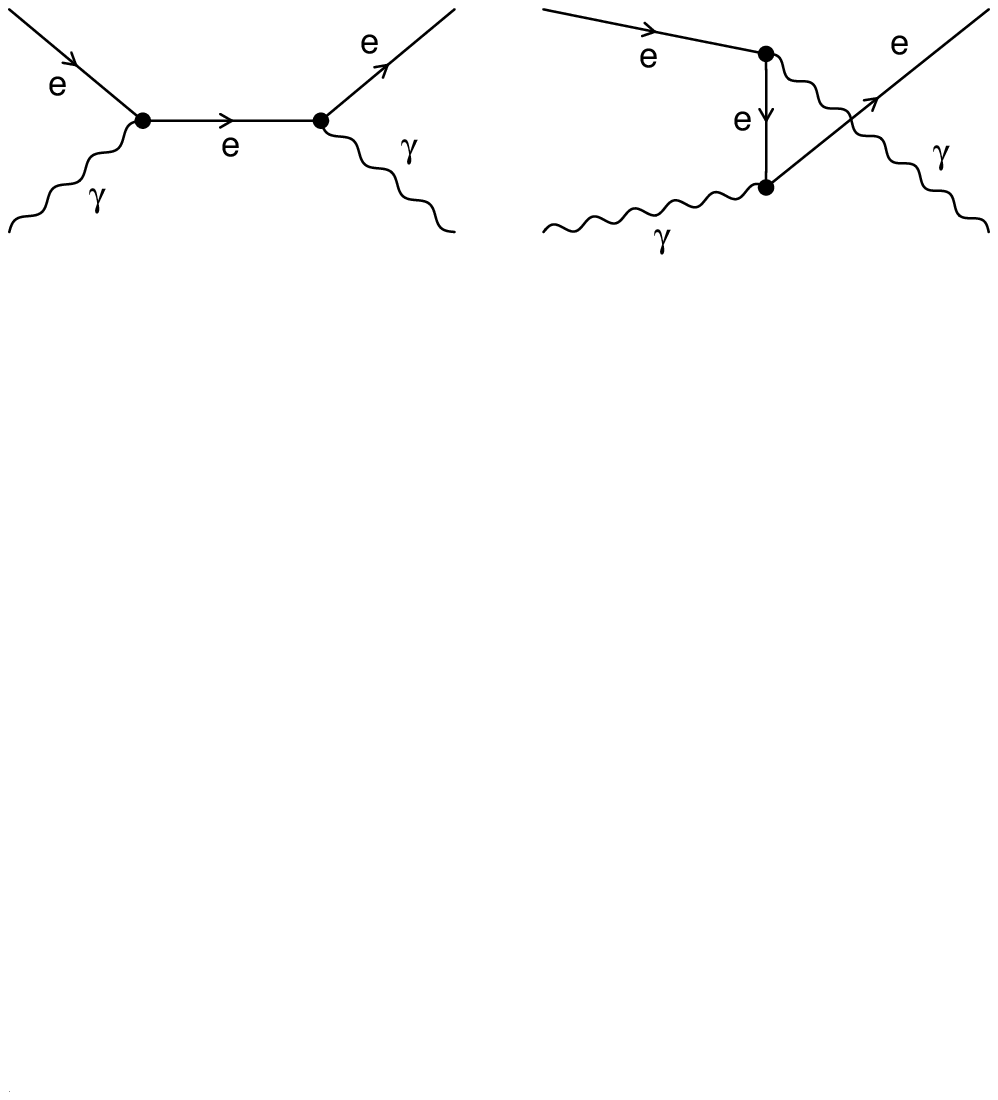}}
\end{picture}
\end{center}
\caption{Tree diagrams for $\Pem\gamma\to\Pem\gamma$.}
\label{egfibdia}
\end{figure}

Since we use the notation and conventions of \cite{egeg,egnwez}
throughout it suffices to repeat here just the most basic definitions.
The helicities of the incoming (outgoing) electron and photon are
denoted by $\sigma_\Pe$ ($\sigma'_\Pe$) and $\lambda_\gamma$
($\lambda'_\gamma$), respectively. The Mandelstam variables are given in
the centre-of-mass system (CMS) by
\beq
s = 4E^{2}, \qquad
t = -4E^2\sin^{2}\frac{\theta}{2}, \qquad
u = -4E^2\cos^{2}\frac{\theta}{2}
\eeq
where $E$ represents the beam energy and $\theta$ the scattering angle
of the electrons (and also the photons). Here we already make use
of the fact that we are interested in energies $E\gg\Me$. Consequently 
we neglect the electron mass whenever possible so that our results are
valid for $s,-t,-u\gg\Me^2$. In this limit the two lowest-order Feynman 
diagrams%
\footnote{All Feynman diagrams of this work have been drawn with the
help of {\it FeynArts} \cite{fa}.}
shown in \reffi{egfibdia} yield the following differential 
cross-sections
\beqar
\left(\frac{d\si}{d\Omega}\right)_{\born} &=& \left\{
\barr{cl}
\disp\alpha^2\frac{1}{(-u)} \quad
& \mbox{for}\quad\sigma_\Pe=\sigma'_\Pe=\pm\frac{1}{2},\,\,
(\lambda_\gamma,\lambda'_\gamma)=(\pm 1,\pm 1), \\[.8em]
\disp\alpha^2\frac{(-u)}{s^2} \quad
& \mbox{for}\quad\sigma_\Pe=\sigma'_\Pe=\pm\frac{1}{2},\,\,
(\lambda_\gamma,\lambda'_\gamma)=(\mp 1,\mp 1), \\[.8em]
0 & \mbox{otherwise.}
\earr\right.
\label{gdiffborn}
\eeqar
Hence both the electron and photon helicities are conserved in lowest
order. Moreover there are two different symmetries: the Born cross-sections 
are invariant under simultaneous reversal of all helicities and 
the transition amplitudes%
\footnote{The squared amplitudes and the cross-sections just differ 
by a trivial (flux) factor $\propto 1/s$ which is not affected by this 
substitution.}
under the interchange
($s\leftrightarrow u, \lambda_\gamma\leftrightarrow -\lambda'_\gamma$).
The former is due to parity conservation, the latter expresses
crossing symmetry.

\section{\sloppy Virtual electroweak and soft-photonic radiative corrections}
\label{virc}

\begin{figure}
\begin{center}
\begin{picture}(16,5.0)
\put(-2.5,-12.5){\includegraphics{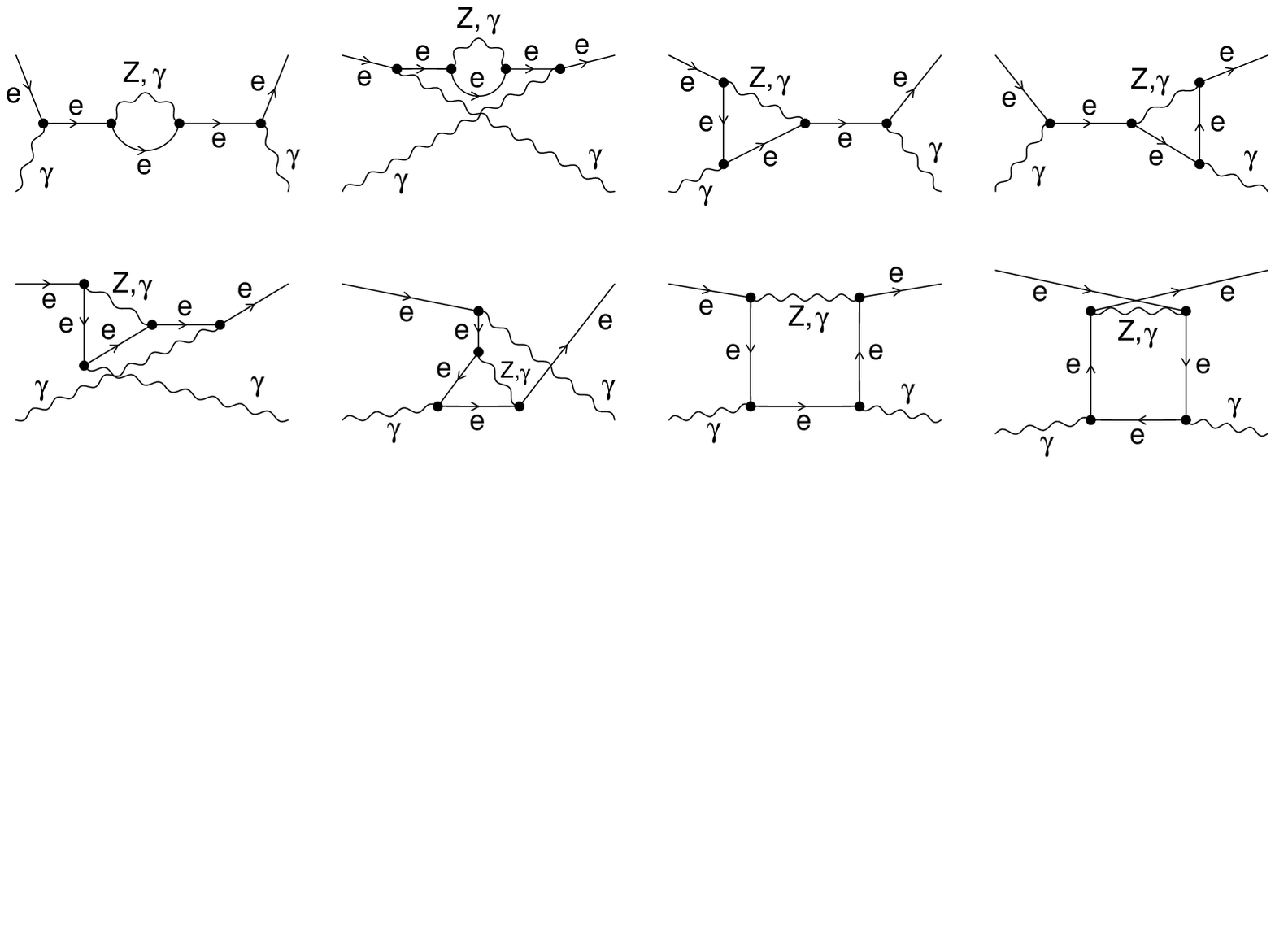}}
\end{picture}
\end{center}
\caption{Feynman diagrams for the exchange of virtual photons and Z bosons.}
\label{qedncdiagrams}
\end{figure}

\begin{figure}
\begin{center}
\begin{picture}(16,7.7)
\put(-2.5,-9.7){\includegraphics{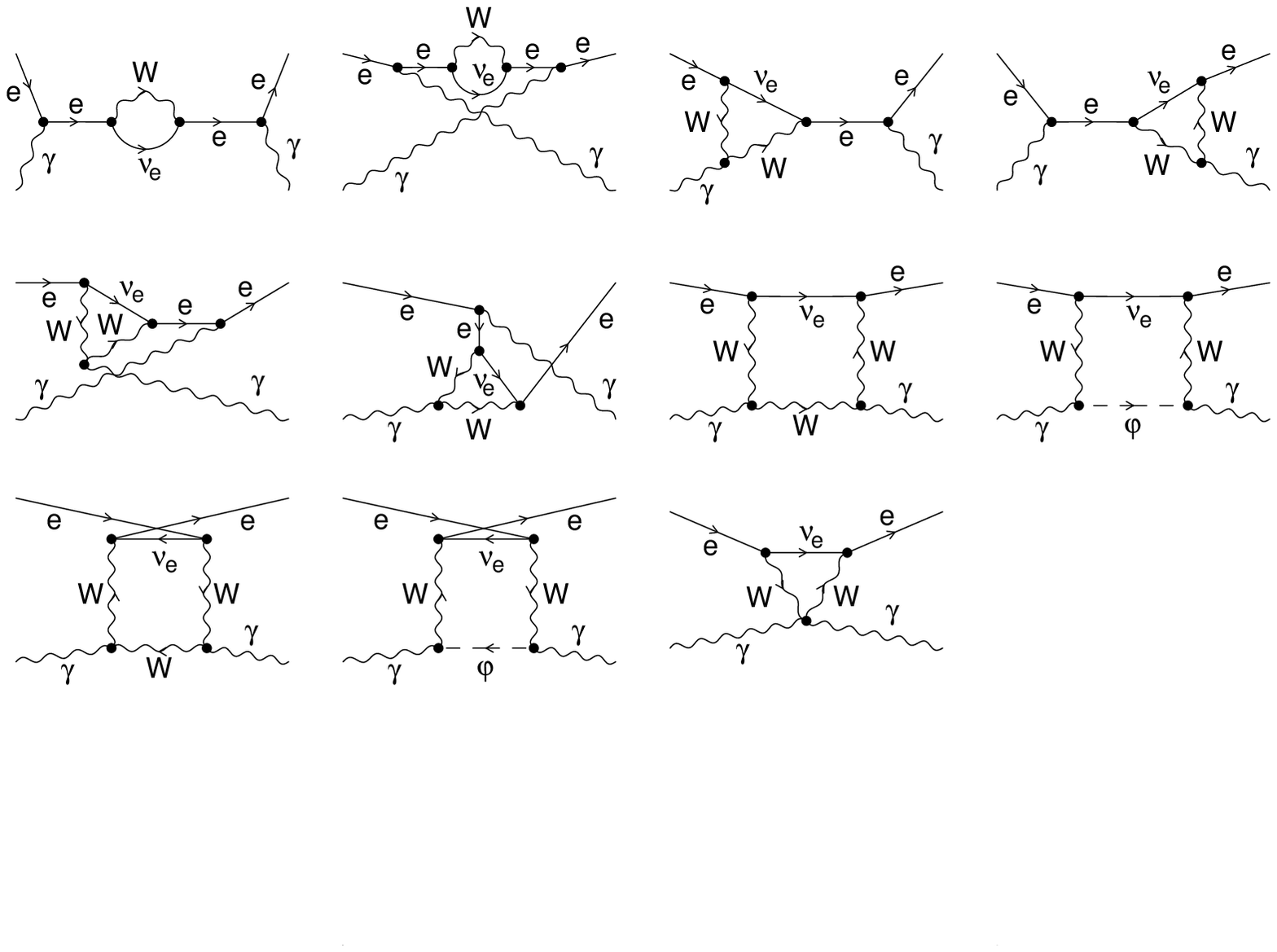}}
\end{picture}
\end{center}
\caption{Feynman diagrams for the exchange of virtual W bosons.}
\label{wdiagrams}
\end{figure}

Analogously to \cite{egeg,egnwez} we calculate the virtual RCs in the
't Hooft-Feynman gauge applying the complete on-shell renormalization
scheme as described in \cite{aoki,hab&mex}, where a complete list of the
renormalization constants can be found. In particular the fields are
normalized in such a way that the residues of all renormalized
propagators are equal to one, i.e.\ self-energy contributions of
external particles drop out. Expanding the squared transition-matrix 
element $\vert\M\vert^2$ for \egeg\ up to $\O(\alpha)$ and taking into
account the soft-photonic bremsstrahlung factor $\de_{\mathrm{SB}}$ 
yields for the differential cross-section
\beqar
\left(\frac{d\si}{d\Omega}\right) &=&
\sum_{\si_\Pe,\si'_\Pe,\lambda_\gamma,\lambda'_\gamma}
\frac{1}{4}(1+2\si_\Pe P_{\Pe})(1+\lambda_{\gamma}P_{\gamma})
\frac{1}{64\pi^{2}s} 
\left[\vert\M_{\mathrm{Born}}\vert^{2}
(1+\de_{\mathrm{SB}})+
2\mbox{Re}\{\M^*_{\mathrm{Born}}\de\M\}\right] \nn \\[.2em]
&=& \left(\frac{d\si}{d\Omega}\right)_{\mathrm{Born}}
(1+\de_{\mathrm{virt}}+\de_{\mathrm{SB}}),
\eeqar
where $P_{\Pe,\gamma}$ denote the degrees of beam polarization.
Following the treatment of \cite{egeg,egnwez} we decompose the virtual 
electroweak RCs, which are summed up in $\de_{\mathrm{virt}}$, into
gauge-invariant subsets
\beq
\de_{\mathrm{virt}} = \delta_{\QED}^{\mathrm{virt}}+
\delta_{\NC}+\delta_{\PW}.
\eeq
$\delta_{\QED}^{\mathrm{virt}}$ and $\delta_{\NC}$ include
all contributions which are due to photon and Z-boson exchange,
respectively. The corresponding Feynman diagrams are shown in
\reffi{qedncdiagrams}. The remaining diagrams, which are shown in
\reffi{wdiagrams}, contain virtual \PW\ bosons and form
$\delta_{\PW}$.

The one-loop RCs vanish for all polarizations corresponding to a Born 
cross-section which is identically zero and vice versa so that we have 
to deal with the polarization combinations $\kappa=\si_\Pe=\si'_\Pe$ and 
$\rho=\lambda_\gamma=\lambda'_\gamma$ for $\delta_{\mathrm{virt}}$ only.%
\footnote{Note that instead of the proper values $\kappa=\pm\frac{1}{2}$
we use the abbreviation $\kappa=\pm$ in subscripts and superscripts to
distinguish right- and left-handed quantities.}
Moreover it turns out that each contribution to the virtual RCs
respects crossing symmetry. Consequently we just have to give 
$\delta^\kappa_{\mathrm{virt}}(\rho=-1)$ and obtain the case
$\rho=+1$ via
\beq
\delta^\kappa_{\mathrm{virt}}(\rho) \;\;=\;\;
\delta^\kappa_{\mathrm{virt}}(-\rho)\Big\vert_{s\leftrightarrow u}. 
\eeq
The following results for $\delta_{\QED}^{\mathrm{virt},\kappa}(\rho)$,
$\delta_{\NC}^\kappa(\rho)$, and $\delta_{\PW}^\kappa(\rho)$ have been
obtained by the application of the computer-algebraic packages 
{\it FeynArts} \cite{fa} and {\it FeynCalc} \cite{fc}. The former
program generates the Feynman diagrams together with the corresponding 
transition-matrix elements for a given scattering process directly from 
the Feynman rules, the latter evaluates the Dirac algebra and reduces 
one-loop tensor integrals to scalar integrals. Finally these 
integrals, which are summarized in \refapp{scalar}, have been calculated 
by hand using the results and methods of \cite{scalar}.

\subsection{Photon exchange}

In order to regularize possible IR divergences we introduce an
infinitesimal mass $\lambda$ for internal photons if necessary.
Inspecting the results for the virtual one-loop QED RCs 
\beqar
\delta_{\QED}^{\mathrm{virt},-}(-) &=& \frac{\alpha}{\pi} 
\left\{
\log\left(\frac{s+\ie}{-\lambda^2}\right)
\left[ 1+\log\left(\frac{\Me^2}{-t}\right) \right]
-\frac{1}{2}\log\left(\frac{-\Me^2}{u+\ie}\right)
+\frac{1}{2}\log^2\left(\frac{-\Me^2}{s+\ie}\right) \right.
\nn\\[.2em]
&& \hspace{-3em} \phantom{\times\Biggr\{}
\left.
+\log\left(\frac{t}{s+\ie}\right)
+\frac{t}{s}\log\left(\frac{t}{u+\ie}\right)
+\frac{t^2}{2s^2}\log^2\left(\frac{t}{u+\ie}\right)
-\frac{3}{2}
+\frac{4s^2+3t^2}{s^2}\zeta(2) \right\}, 
\nn\\[.2em]
\delta_{\QED}^{\mathrm{virt},+}(\rho) &=&
\delta_{\QED}^{\mathrm{virt},-}(-\rho),
\label{qedrc}
\eeqar
we find that it is entirely formed by logarithms.

\subsection{Z-boson exchange}

In \cite{egeg,egnwez} it has been pointed out that the set of all 
Feynman diagrams which contain internal Z bosons form a gauge-invariant subset.
The result for these contributions, which have been called {\it neutral 
current corrections}, reads
\beqar
\delta_{\NC}^-(-) &=& \frac{\alpha}{\pi}
\left(g_{\Pe\Pe\PZ}^-\right)^2 
\left\{
\left(1-\frac{\MZ^2}{u}\right) 
\left(\frac{3}{2}+\frac{u}{s}+\frac{\MZ^2}{2u}-\frac{\MZ^2}{s}\right) 
\log\left(1-\frac{u+\ie}{\MZ^2}\right) 
-\frac{u}{s}\log\left(\frac{-t}{\MZ^2}\right)
\right. \nn\\[.2em]
&& \hspace{-4.5em} \phantom{\times\Biggr\{}
\left.
+\frac{(t+\MZ^2)^2}{s^2} \left[ \zeta(2)
-\log\left(\frac{-t}{\MZ^2}\right)
\log\left(1-\frac{u+\ie}{\MZ^2}\right) 
-\Li\left(\frac{u+\ie}{\MZ^2}\right) 
-\Li\left(1+\frac{t}{\MZ^2}\right) 
\right] \right. \nn\\[.2em]
&& \hspace{-4.5em} \phantom{\times\Biggr\{}
\left.
-\frac{(s-\MZ^2)^2}{s^2} \left[
\log\left(\frac{-t}{\MZ^2}\right)
\log\left(1-\frac{s+\ie}{\MZ^2}\right) 
+\Li\left(\frac{s+\ie}{\MZ^2}\right) \right] 
+ \frac{\MZ^2}{s} - \frac{\MZ^2}{2u} - \frac{5}{4} 
\right\},
\nn\\[.2em]
\delta_{\NC}^+(\rho) &=&
\delta_{\NC}^-(-\rho)\;
(g_{\Pe\Pe\PZ}^+/g_{\Pe\Pe\PZ}^-)^2,
\label{ncrc}
\eeqar
where the electron-Z couplings are abbreviated by
\beq
g_{\Pe\Pe\PZ}^+=\frac{\sw}{\cw}, \qquad 
g_{\Pe\Pe\PZ}^-=\frac{\sw}{\cw} - \frac{1}{2\sw\cw}.
\eeq
Note that the individual scalar integrals (see \refapp{scalar}) contain
mass-singular logarithms of the electron which drop out in 
$\delta_{\NC}$.

\subsection{W-boson exchange}
\label{wvirc}

In contrast to the virtual QED and neutral current RCs the analytical
results for the contributions caused by W-boson exchange do not simplify
after inserting the explicit expressions for the scalar integrals.
Therefore we introduce some abbreviations for these integrals which are
explicitly given in \refapp{scalar}
\beqar
\delta_{\PW}^-(-) &=& \frac{\alpha}{4\pi\sw^2}  \nn\\[.2em]
&& \hspace{-5em} \times \left\{
\frac{1}{2} + \frac{2u}{s}(1-B_{t\PW\PW})
+\frac{u(2u-3s)+\MW^2(2u-s)}{su}B_{u0\PW}
+\frac{(u-s)(u+2\MW^2)}{s^2}utD_{ut}
\right. \nn\\[.2em]
&& \hspace{-5em} \phantom{\times\Biggr\{}
\left.
+\frac{u(s-u-2\MW^2)}{s^2}
\left[t(C_{t\PW\PW}+C_{t\PW\PW\PW})+2uC_{u\PW\PW}
-t\MW^2D_{ut}\right]
-\frac{(s-\MW^2)^2}{s^2} 
\right. \nn\\[.2em]
&& \hspace{-5em} \phantom{\times\Biggr\{}
\times\left[
2(sC_{s\PW\PW}+uC_{u\PW\PW}+tC_{t\PW\PW\PW})
+(s^2-su-t\MW^2)D_{st}
+(u^2-us-t\MW^2)D_{ut} \right]
\Biggl\}, \nn\\[.2em]
\delta_{\PW}^+(\rho) &=& 0.
\label{wrc}
\eeqar

Recall that the comparably short results presented in (\ref{qedrc}), 
(\ref{ncrc}), and (\ref{wrc}) have to be compared
with lengthy and untransparent formulae of \cite{egnwez}.

\subsection{Soft bremsstrahlung}

The soft-photonic bremsstrahlung factor $\de_{\mathrm{SB}}$, which is
universal for all polarizations, can be simply cited from \cite{egnwez}
\beqar
\de_{\mathrm{SB}} & = & -\frac{\alpha}{\pi}\Biggr\{
\log\left(\frac{4\Delta E^{2}}{\lambda^{2}}\right)
\left[1+\log\left(\frac{\Me^{2}}{-t}\right)\right]
+2\zeta(2)
+\Li\left(\frac{-u}{t}\right)
+\log\left(\frac{\Me^{2}}{s}\right)
\nn \\[.4em]
&& \phantom{\frac{\alpha}{2\pi}\Big\{}
+\frac{1}{2}\log^{2}\left(\frac{\Me^{2}}{s}\right) \Biggr\}.
\label{sbrc}
\eeqar
Here $\Delta E\ll E$ represents the soft-photon cut-off. 
In \cite{egeg,egnwez} the QED corrections have been defined by
\beq
\delta_{\QED} = \delta_{\QED}^{\mathrm{virt}}+\de_{\mathrm{SB}}.
\eeq
Inspecting $\delta_{\QED}^{\mathrm{virt}}$ and $\de_{\mathrm{SB}}$ one
easily verifies that both the IR-divergent terms ($\log\lambda$) as 
well as the {\it Sudakov logarithms} ($\log^2\Me$) cancel in 
$\delta_{\QED}$, and that the remaining mass-singular logarithms 
($\log\Me$) agree with the ones predicted by structure function
methods \cite{cdstruct}.

\section{Hard-photonic bremsstrahlung -- double Compton scattering}
\label{hbrc}

In this section we deal with double Compton scattering, i.e.\ with the
reaction
$$\Pem(p_{\Pe},\sigma_\Pe) + \gamma(k_{\gamma},\lambda_{\gamma})
\to\Pem(p^{\prime}_{\Pe},\sigma^{\prime}_{\Pe}) + \gamma(k_1,\lambda_1) +
\gamma(k_2,\lambda_2).$$
Using throughout the conventions and calculational techniques of
\cite{egnwgezg}, where the radiative processes \egnwgezg\ have been
discussed, we are able to present the results in a very compact
manner.

\subsection{Non-collinear photon emission}

We first restrict our treatment to non-collinearly emitted photons so
that we can neglect the electron mass for beam energies $E\gg\Me$. Owing
to $\Me=0$ the electron helicity is conserved and we can define 
$\kappa=\sigma_\Pe=\sigma^{\prime}_\Pe$. The originally $2^4=16$
independent amplitudes $\M^\kappa(\lambda_\gamma,\lambda_1,\lambda_2)$
are reduced to four independent polarization combinations by the
following discrete symmetries:
\beqar
\mbox{Bose symmetry:} &\;\;\;&
\rlap{$\M^\kappa(\lambda_\gamma,\lambda_1,\lambda_2) =
\M^\kappa(\lambda_\gamma,\lambda_2,\lambda_1) $}
\hspace{19em}\mbox{with } 1\leftrightarrow 2, \hspace{2em}
\\[.4em]
\mbox{crossing symmetry:} &&
\rlap{$\M^\kappa(\lambda_\gamma,\lambda_1,\lambda_2) =
\M^\kappa(-\lambda_1,-\lambda_\gamma,\lambda_2) $}
\hspace{19em}\mbox{with } \gamma\leftrightarrow 1,
\nn\\
&&
\rlap{$\phantom{\M^\kappa(\lambda_\gamma,\lambda_1,\lambda_2)}
= \M^\kappa(-\lambda_2,\lambda_1,-\lambda_\gamma) $}
\hspace{19em}\mbox{with } \gamma\leftrightarrow 2,
\\[.4em]
\mbox{parity conservation:} &&
\M^\kappa(\lambda_\gamma,\lambda_1,\lambda_2) =
-\M^{-\kappa}(-\lambda_\gamma,-\lambda_1,-\lambda_2)^*.
\eeqar

\begin{figure}
\begin{center}
\begin{picture}(14,6.5)
\put(-2.5,-9.5){\includegraphics{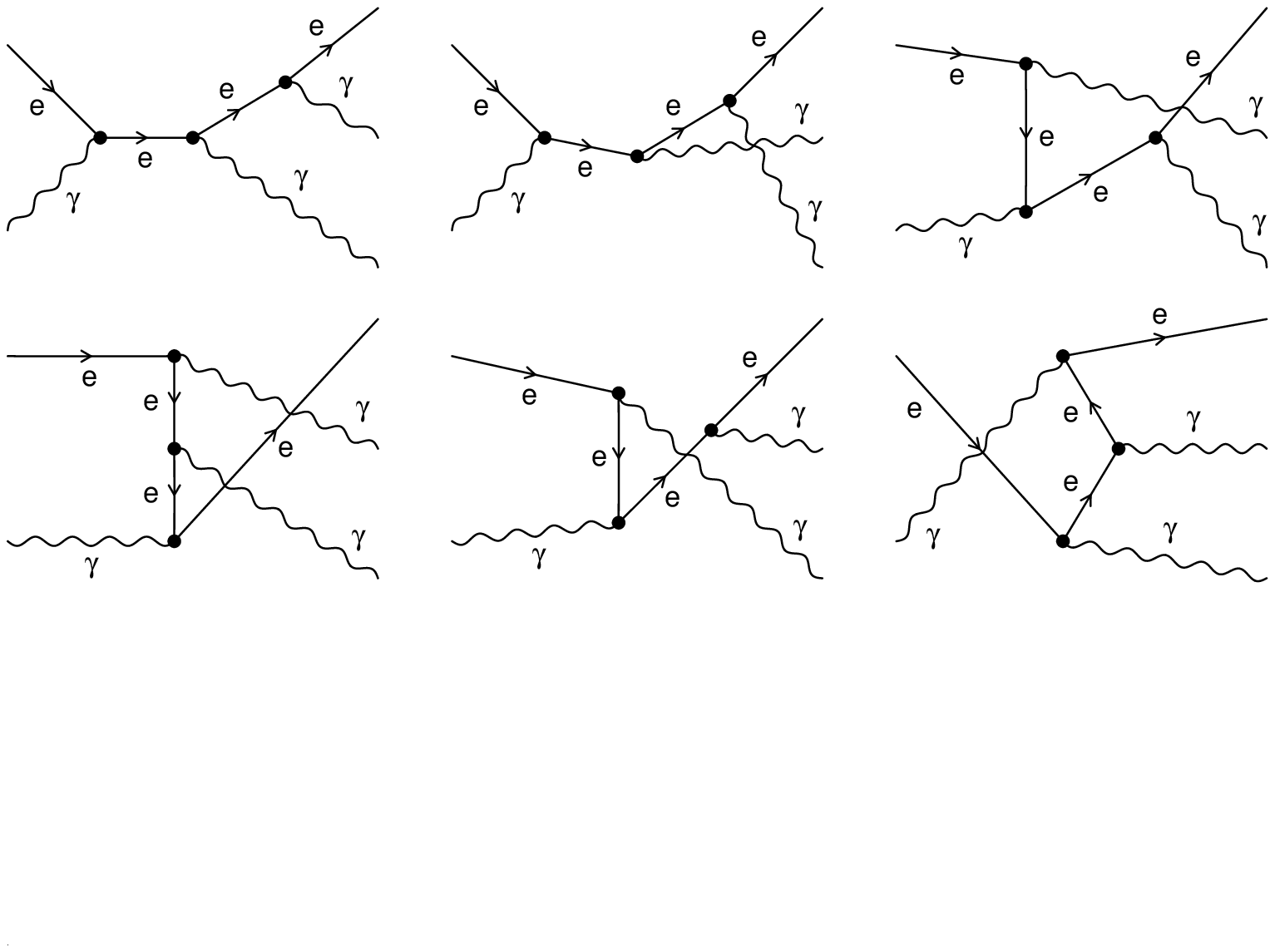}}
\end{picture}
\end{center}
\caption{Tree diagrams for $\Pem\gamma\to\Pem\gamma\gamma$.}
\label{brems}
\end{figure}
Figure \ref{brems} shows the tree diagrams describing double Compton
scattering in lowest order. In terms of Weyl-van der Waerden spinor products
the independent transition-matrix elements for $\kappa=-1/2$ read
\beqar
\M^-(+1,-1,-1) &=& \M^-(-1,+1,+1) = 0,  \nn\\[.3em]
\M^-(+1,+1,+1) &=& 2\sqrt{2}ie^3
\frac{\langle p_\Pe k_\gamma\rangle^*(\langle p'_\Pe k_\gamma\rangle)^2
\langle p_\Pe p'_\Pe\rangle}
{\langle p_\Pe k_\gamma\rangle
\langle p'_\Pe k_1\rangle^*\langle p_\Pe k_1\rangle
\langle p'_\Pe k_2\rangle^*\langle p_\Pe k_2\rangle},
\nn\\[.3em]
\M^-(-1,-1,-1) &=& 2\sqrt{2}ie^3
\frac{\langle p'_\Pe k_\gamma\rangle(\langle p_\Pe k_\gamma\rangle^*)^2
\langle p_\Pe p'_\Pe\rangle^*}
{\langle p'_\Pe k_\gamma\rangle^*
\langle p'_\Pe k_1\rangle^*\langle p_\Pe k_1\rangle
\langle p'_\Pe k_2\rangle^*\langle p_\Pe k_2\rangle},
\label{eggamp}
\eeqar
where the spinorial products $\langle\phi\psi\rangle$ are defined
like in \cite{egnwgezg}.
Moreover it is instructive to express the invariant spinor products of
(\ref{eggamp}) in terms of particle energies ($E,E'_\Pe,E_1,E_2$) and
scattering angles for the squared matrix elements:
\beqar
\vert\M^-(+1,+1,+1)\vert^2 &\,=\,& 2e^6
\frac{\disp EE'_\Pe\cos^4\frac{\theta'_\Pe}{2}\sin^2\frac{\theta'_\Pe}{2}}
{\disp E_1^2 E_2^2 \sin^2\frac{\theta_1}{2}   \sin^2\frac{\theta_2}{2}
		   \sin^2\frac{\alpha_1}{2}   \sin^2\frac{\alpha_2}{2}},
\nn\\[.3em]
\vert\M^-(-1,-1,-1)\vert^2 &\,=\,& 2e^6
\frac{\disp E^3          \sin^2\frac{\theta'_\Pe}{2}}
{\disp E'_\Pe E_1^2 E_2^2\sin^2\frac{\theta_1}{2}   \sin^2\frac{\theta_2}{2}
                         \sin^2\frac{\alpha_1}{2}   \sin^2\frac{\alpha_2}{2}},
\label{eggamp2}
\eeqar
with $\theta'_\Pe,\theta_i$ $(i=1,2)$ denoting the polar angles of the
corresponding particles with respect to the direction of the incoming 
electron and $\alpha_i=\angle({\bf k}_i,{\bf p}'_\Pe)$.
From (\ref{eggamp2}) we can read for instance the structure of the
IR and collinear poles in $E_i$, $\theta_i$ and $\alpha_i$,
respectively, which is different for the single polarization combinations. 
On the other hand all amplitudes contain the global factor
$\sin^2\frac{\theta'_\Pe}{2}$ so that non-collinear double Compton 
scattering is suppressed for forward scattering of the electron 
$(\theta'_\Pe\to 0)$.

\subsection{Collinear photon emission -- finite-mass effects}
\label{collpho}

One possible way to include the finite-mass effects of the electrons in
the case of collinear photon emission has been proposed in \cite{finitemass}.
This method is based on the introduction of {\it mass-effective
factors}
\beqar
f^{\mathrm{(ini/fin)}}_+(\xi,\varepsilon,\theta) &\,=\,& \left(
\frac{4\varepsilon^2\sin^2\frac{\theta}{2}}
{4\varepsilon^2\sin^2\frac{\theta}{2}+\Me^2} \right)^2, \nn\\[.3em]
f^{\mathrm{(ini/fin)}}_-(\xi,\varepsilon,\theta) &\,=\,& 
\frac{\xi^2}{\xi^2\mp 2\xi+2}
\frac{4\varepsilon^2\Me^2\sin^2\frac{\theta}{2}}
{\left(4\varepsilon^2\sin^2\frac{\theta}{2}+\Me^2\right)^2}.
\eeqar
for each initial/final-state electron of energy $\varepsilon\gg\Me$ 
emitting a photon of 
energy $\xi\varepsilon$. Note that the functions $f_+$ and $f_-$ are 
different from 1 and 0, respectively, only for emission angles 
$\theta\sim\Me /\varepsilon$. In this collinear region the
singular behaviour of $\vert\M\vert^2$ for $\Me=0$ is replaced by the
correct asymptotics via
\beqar
\sum_{\lambda_1,\lambda_2=\pm 1}
\vert 
\M(\sigma_\Pe,\sigma'_\Pe,\lambda_\gamma,\lambda_1,\lambda_2) 
\vert^2 
\Big\vert_{\Me\ne 0} & = & 
\sum_{\kappa_1,\kappa'_1=\pm 1\atop\kappa_2,\kappa'_2=\pm 1}
\left( \prod_{i=1,2}
f^{\mathrm{(ini)}}_{\kappa_i}(x_i,E,\theta_i) \, 
f^{\mathrm{(fin)}}_{\kappa'_i}(x'_i,E'_\Pe,\alpha_i) \right) 
\nn\\
&& \hspace{-2em}\times 
\sum_{\lambda_1,\lambda_2=\pm 1}
\vert \M(\kappa_1\kappa_2\sigma_\Pe,\kappa'_1\kappa'_2\sigma'_\Pe,
\lambda_\gamma,\lambda_1,\lambda_2) 
\vert^2 \Big\vert_{\Me =0} 
\nn\\[.4em]
&& \hspace{-8em} \mbox{with:}\qquad 
x_i = E_i/E,\quad
x'_i = E_i/E'_\Pe,\quad
i=1,2.
\label{masseff}
\eeqar
In this context we should mention that (\ref{masseff}) is not valid for
double collinear photon emission. However, this situation can not occur
if we impose an angular cut 
$\theta^{\mathrm{for}}_\Pe<\theta'_\Pe<\theta^{\mathrm{back}}_\Pe$ on the
electron scattering angle. Consequently in this angular region the spins 
of the electrons can not be flipped twice, in other words there will be
no contributions to the sum in (\ref{masseff}) if more than one of the
$\kappa_{1,2}^{(\prime)}$ are equal to $-1$.

Following this method of mass-effective factors collinearly emitted
photons are included by suitable modifications of the squared amplitude
so that the photon phase space is not restricted. In \refse{collint} we
will describe an alternative method where the collinear regions of
the phase space are treated separately.

\subsection{Phase-space integration}

\begin{figure}
\begin{center}
\begin{picture}(14,8.4)
\put(-5.5,-9.3){\includegraphics{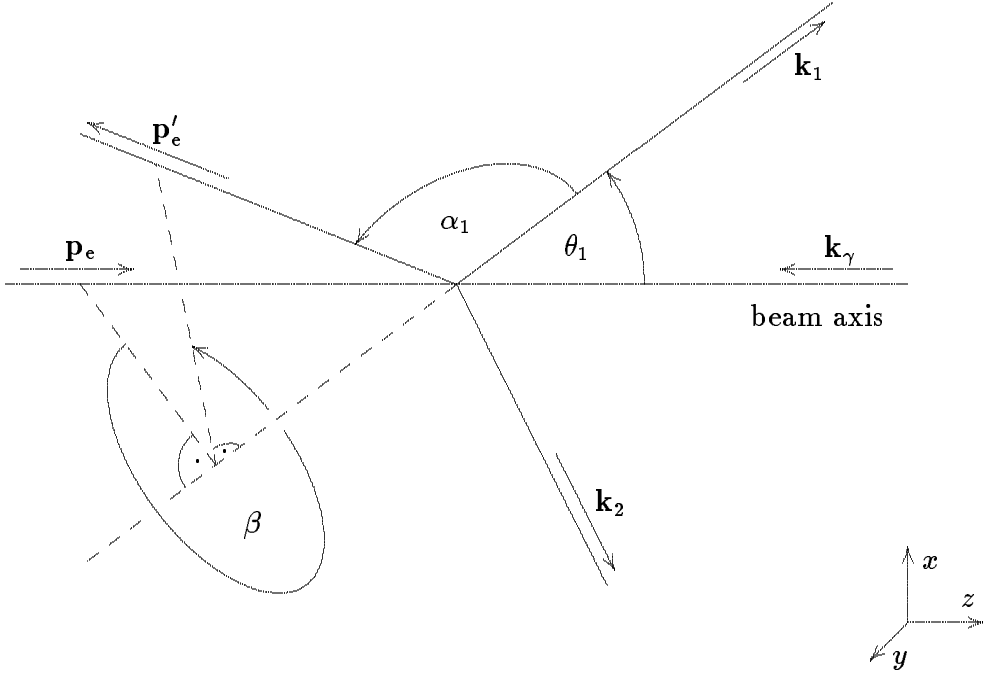}}
\end{picture}
\end{center}
\caption{Particle kinematics for $\phi_1=0$.}
\label{kinema}
\end{figure}

In the CMS the total cross-section for double Compton scattering is
given by
\beq
\sigma_{\mathrm{tot}} = \frac{1}{8E^2}
\sum_{\sigma_\Pe,\lambda_\gamma\atop\sigma'_\Pe,\lambda_1,\lambda_2}
\frac{1}{2}(1+2P_\Pe\sigma_\Pe)
\frac{1}{2}(1+P_\gamma\lambda_\gamma) \int_\Gamma d\Gamma\,
\vert \M(\sigma_\Pe,\sigma'_\Pe,\lambda_\gamma,\lambda_1,\lambda_2)
\vert^2.
\label{sigegegg}
\eeq
Figure \ref{kinema} illustrates the particle kinematics where ${\bf k}_1$
is oriented into the $x$-$z$ plane making use of rotational invariance
around the beam axis. $\alpha_1$ and $\beta$ denote the polar and 
azimuthal angle of ${\bf p}^{\prime}_\Pe$ relative to ${\bf k}_1$, 
respectively.
\beq
{\bf p}^{\prime}_\Pe = E'_\Pe
\pmatrix{\cos\theta_1 & 0 & \sin\theta_1 \cr 0 & 1 & 0 \cr
-\sin\theta_1 & 0 & \cos\theta_1}
\pmatrix{ \cos\beta\sin\alpha_1 \cr \sin\beta\sin\alpha_1 
\cr \cos\alpha_1}
\qquad \mbox{for}\quad \phi_1=0.
\eeq
Owing to the fact that the two emitted photons are identical
particles we have to symmetrize the polarized
cross-sections and to apply a factor $1/2$ when integrating over the
photon phase space. Instead of this factor $1/2$ we prefer to impose the
constraint
\beq
E_1\sin\frac{\theta_1}{2}\sin\frac{\alpha_1}{2}<
E_2\sin\frac{\theta_2}{2}\sin\frac{\alpha_2}{2}
\label{symcut}
\eeq
which cuts the phase space in half in a symmetric way. Moreover
(\ref{symcut}) ensures that only photon `1' may become an `IR' or
`collinear photon'. For this reason we choose the following 
parametrization for the phase space
\beqar
\int_\Gamma d\Gamma &=&
\int\frac{d^3 {\bf p}'_\Pe}{(2\pi)^32p^{\prime0}_\Pe} \,
\int\frac{d^3 {\bf k}_1}{(2\pi)^32k^0_1} \,
\int\frac{d^3 {\bf k}_2}{(2\pi)^32k^0_2} \,
(2\pi)^4\delta^{(4)}(p_\Pe+k_\gamma-p'_\Pe-k_1-k_2) \nn\\
&=& \frac{1}{8(2\pi)^5} \int dE_1\,d\Omega_1\,d\beta\,
d\cos\alpha_1\,\left\vert\frac{\partial E'_\Pe}{\partial\cos\alpha_1}
\right\vert,
\label{phasespace}\eeqar
with
\beq
\frac{\partial E'_\Pe}{\partial\cos\alpha_1} =
-\frac{E_1E(E-E_1)}{2\left(E-E_1\sin^2\frac{\alpha_1}{2}\right)^2},
\eeq
which singles out both collinear singularities ($\theta_1\to 0,
\alpha_1\to 0$) as well as the IR pole ($E_1\to 0$). Furthermore these 
poles of the integrands have been transformed away by appropriate
transformations since numerical integration works best for flat
functions. We have applied the well-known Monte Carlo routine 
{\it VEGAS} \cite{vegas}.

Apart from the different cuts the integration boundary is given by
\beq
0\le\theta_1\le\pi, \quad 0\le\beta\le 2\pi, \quad
0\le\alpha_1\le\pi, \quad 0\le E_1\le E,
\eeq
while the integration over $\phi_1$ trivially yields a factor $2\pi$.
In order to resolve the electron and at least one photon in the final
state we restrict the phase space by
\beqar
\theta^{\mathrm{for}}_\Pe<\theta'_\Pe<\theta^{\mathrm{back}}_\Pe,
& \quad & E'_\Pe>\delta_\Pe E, \nn\\[.2em]
\theta^{\mathrm{for}}_\gamma<\theta_i<\theta^{\mathrm{back}}_\gamma,
&& E_i>\delta_\gamma E
\qquad\mbox{for}\quad i=1\quad\mbox{or}\quad i=2.
\label{eggcuts}
\eeqar

\subsection{Integration over the collinear regions}
\label{collint}

In \cite{egnwgezg} the integration over the collinear regions has
been carried out analytically both for initial and final-state
radiation. Here we can directly make use of the results given there and
obtain for collinear initial-state radiation, i.e.\ for emission angles
$\theta_1<\Delta\theta_1\ll 1$, the following integral representation:
\beqar
\sigma^{\mathrm{coll,in}}_{\mathrm{tot}}(E) &=& \frac{\alpha}{2\pi}
\Biggl\{\,(L_\Pe-1)\int_{x_{\mathrm{min}}}^{x_{\mathrm{max}}}dx\,
\frac{x^2-2x+2}{x}\,
\hat\sigma^{(x)}_{\mathrm{tot}}\big(\sqrt{1-x}E\big) \nn\\
&& \phantom{\frac{\alpha}{2\pi}\Biggl\{\,(L_\Pe-1)}
+\int_{x_{\mathrm{min}}}^{x_{\mathrm{max}}}dx\,x\,
\hat\sigma^{(x)}_{\mathrm{tot}}\big(\sqrt{1-x}E\big)
\Big\vert_{P_\Pe\to -P_\Pe}\,\Biggl\},
\label{incollint}
\eeqar
with
\beq
L_\Pe = \log\left(\frac{\Delta\theta_1^2E^2}{\Me^2}\right).
\eeq
Of course the mass-singular logarithm $\log\Me$ of (\ref{incollint}) 
agrees with the one predicted in \cite{cdstruct}. While the lower
limit for $x$ is simply set by the soft-photon cut $\Delta E\ll E$ 
the determination of the maximal $x$ is more complicated. 
The various energy and angular cuts (\ref{eggcuts}) lead to
\beq
x_{\mathrm{min}} = \frac{\Delta E}{E}, \qquad
x_{\mathrm{max}} = \mathrm{min}\left\{
\frac{1-\delta_\Pe}{1-\delta_\Pe\sin^2\frac{\theta_\Pe^{\mathrm{back}}}{2}},
\frac{1-\delta_\gamma}
{1-\delta_\gamma\sin^2\frac{\theta_\gamma^{\mathrm{back}}}{2}} \right\}.
\eeq
In (\ref{incollint}) $\hat\sigma^{(x)}_{\mathrm{tot}}\big(\sqrt{1-x}E\big)$
represents the Born cross-section of the `hard process' \egeg\ which is 
transformed into the boosted CMS of $(1-x)p_\Pe$ and $k_\gamma$ and
calculated by integrating the differential cross-sections (\ref{gdiffborn})
with the beam energy $E^{(x)}=\sqrt{1-x}E$ over the scattering angle
$\theta^{(x)}$ in the boosted system. $\theta^{(x)}$ is related to 
$\theta'_\Pe$ and $\theta_2$ by
\beq
\cos\theta^{(x)} 
\;=\; \frac{2\cos\theta'_\Pe+x(1-\cos\theta'_\Pe)}{2-x(1-\cos\theta'_\Pe)}
\;=\;-\frac{2\cos\theta_2+x(1-\cos\theta_2)}{2-x(1-\cos\theta_2)},
\eeq
which transforms $\theta_\Pe^{\mathrm{for/back}}$ and 
$\theta_\gamma^{\mathrm{for/back}}$ into the ($x$-dependent) angular
cuts $\theta_\Pe^{(x),\mathrm{for/back}}$ and
$\theta_\gamma^{(x),\mathrm{for/back}}$ in the boosted system, 
respectively. However, the cut-offs $\delta_\Pe$ and $\delta_\gamma$ 
for the particle energies influence the angular range $I^{(x)}$ for 
$\theta^{(x)}$, too. Defining
\beqar
\cos\hat\theta_\Pe^{(x)}\;= \phantom{-}
\frac{2}{x}\left(1-\frac{x}{2}-\delta_\Pe\right) 
&\qquad& \mbox{for} \quad x>1-\delta_\Pe,
\nn\\[.2em]
\cos\hat\theta_\gamma^{(x)}\;=
-\frac{2}{x}\left(1-\frac{x}{2}-\delta_\gamma\right) 
&\qquad& \mbox{for} \quad x>1-\delta_\gamma,
\eeqar
$I^{(x)}$ is given by
\beq
I^{(x)} = \left(\;
\mathrm{max}\left\{ \theta_\Pe^{(x),\mathrm{for}},
\hat\theta_\Pe^{(x)},\theta_\gamma^{(x),\mathrm{back}} \right\},\;
\mathrm{min}\left\{ \theta_\Pe^{(x),\mathrm{back}},
\theta_\gamma^{(x),\mathrm{for}},\hat\theta_\gamma^{(x)} \right\}
\;\right).
\eeq

The case of collinear final-state radiation, which corresponds to an
integration over $\alpha_1<\Delta\alpha_1\ll 1$,
is less involved yielding
\beqar
\sigma_{\mathrm{tot}}^{\mathrm{coll,fin}}(\sigma'_\Pe,E) &=& 
\frac{\alpha}{2\pi}\Biggl\{
\hat\sigma_{\mathrm{tot}}(\sigma'_\Pe,E)\Biggl[
\frac{1}{2}(1-\delta_\Pe)(5+\delta_\Pe)+
\delta_\Pe(2+\delta_\Pe)\log\delta_\Pe -
4\Li(1-\delta_\Pe)
\nn\\[.2em]
&& \hspace{2.9em}
+(\hat L_\Pe-1)\left( -2\log\left(\frac{\Delta E}{E}\right)
-\frac{1}{2}(1-\delta_\Pe)(3+\delta_\Pe)+2\log(1-\delta_\Pe) \right)\Biggr]
\nn\\[.2em]
&&\hspace{2em}
+\hat\sigma_{\mathrm{tot}}(-\sigma'_\Pe,E)\frac{1}{2}(1-\delta_\Pe)^2\Biggr\},
\label{fincollint}
\eeqar
with
\beq
\hat L_\Pe = \log\left(\frac{\Delta\alpha_1^2E^2}{\Me^2}\right).
\eeq
Here $\hat\sigma_{\mathrm{tot}}(\sigma'_\Pe,E)$ represents the Born
cross-section for \egeg\ in the original CMS rendering the introduction
of angular cuts trivial.

\section{Results and discussion}
\label{results}

\subsection{Photonic corrections to \protect\egeg}

Since we adopt the input parameters from \cite{egeg,egnwez,egnwgezg} for
numerical evaluations all results fit properly to the ones given there.
The investigation of hard-bremsstrahlung corrections to polarized
Compton scattering is complicated by the fact that both emitted 
photons may become `IR' or `collinear photons'. Although all given
analytical results are sufficient for a discussion of completely
polarized final states here we sum over both helicities 
$\lambda_i=\pm 1$ $(i=1,2)$ and integrate over all singular regions of 
the phase space. 
In accordance with (\ref{eggcuts}) the angular regions are restricted by
$\Delta\theta_\Pe<\theta'_\Pe<180^\circ-\Delta\theta_\Pe$ and
$\Delta\theta_\gamma<\theta_i<180^\circ-\Delta\theta_\gamma$ for $i=1$
or $i=2$. The energy cuts 
$\Delta E_\gamma$ and $\Delta E'_\Pe$ are
both adjusted to the beam energy $E$ according to $\Delta E_\gamma=
\delta_\gamma E$ and $\Delta E'_\Pe=\delta_\Pe E$ with the sample
values $\delta_\gamma=\delta_\Pe=0.2$. 

\begin{figure}[p]
\begin{center}
\begin{picture}(15,8.5)
\put(-.2,-.5){\includegraphics{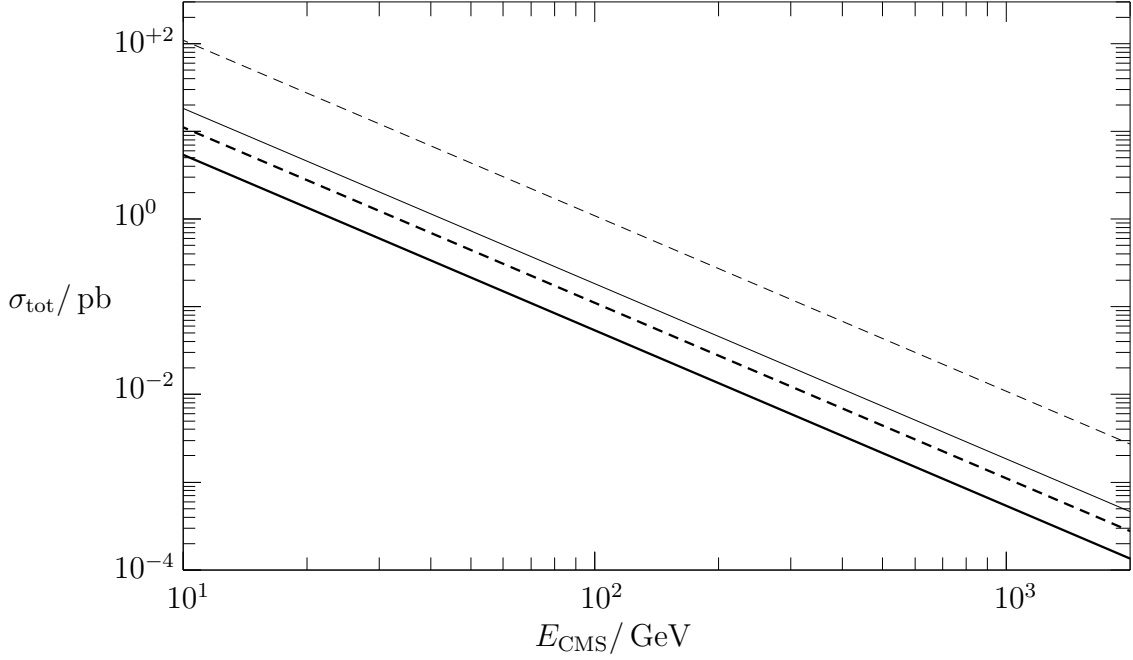}}
\put(0,4.5){$\si_{\mathrm{tot}}/\pba$}
\put(7,0){$E_{\mathrm{CMS}}/\GeV$}
\put(2.1,.5){$10^{1}$}
\put(7.6,.5){$10^{2}$}
\put(13.1,.5){$10^{3}$}
\put(1.4,7.9){$10^{+2}$}
\put(1.4,5.6){$10^{0}$}
\put(1.4,3.3){$10^{-2}$}
\put(1.4,0.9){$10^{-4}$}
\end{picture}
\end{center}
\caption{Cross-sections for helicity-changing channels
($\kappa=\sigma_\Pe=-\sigma'_\Pe$) and polarizations 
$(\kappa,\lambda_{\gamma})$
integrated with $\Delta\theta_\gamma=\Delta\theta_\Pe=20^\circ$ (thick
curves) and $\Delta\theta_\gamma=0^\circ$, $\Delta\theta_\Pe=1^\circ$:
----- $(\mp\frac{1}{2},-1)$, \, $---$ 
$(\mp\frac{1}{2},+1)$.}
\label{eggschang}
\end{figure}
\begin{figure}[p]
\begin{center}
\begin{picture}(15,8.5)
\put(-.2,-.5){\includegraphics{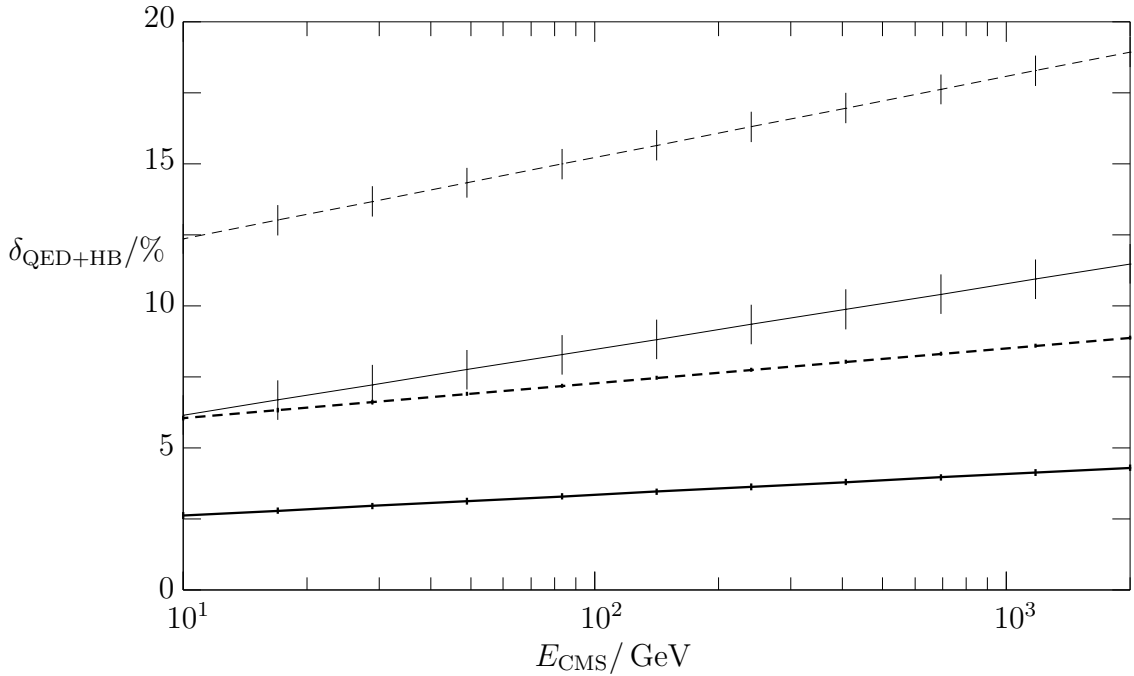}}
\put(0,5.4){$\delta_{\mathrm{QED+HB}}/\%$}
\put(7,0){$E_{\mathrm{CMS}}/\GeV$}
\put(2.1,.5){$10^{1}$}
\put(7.6,.5){$10^{2}$}
\put(13.1,.5){$10^{3}$}
\put(1.8,8.5){$20$}
\put(1.8,6.6){$15$}
\put(1.8,4.7){$10$}
\put(2.0,2.8){$5$}
\put(2.0,0.9){$0$}
\end{picture}
\end{center}
\caption{Photonic
corrections to the integrated
cross-section for different polarization combinations
$(\kappa,\lambda_{\gamma})$ with $\kappa=\sigma_\Pe=\sigma'_\Pe$ (Same
signature as \protect\reffi{eggschang}).}
\label{eggqm}
\end{figure}
Figure \ref{eggschang} shows the total cross-sections for the helicity-changing
channels $(\kappa=\sigma_\Pe=-\sigma'_\Pe)$ for the two different angular cuts
$\Delta\theta_\gamma=\Delta\theta_\Pe=20^\circ$ and
$\Delta\theta_\gamma=0^\circ$, $\Delta\theta_\Pe=1^\circ$. As already
observed for \egnwgezg\ \cite{egnwgezg} these contributions are entirely 
due to collinear photon emission and can be calculated via (\ref{incollint}) 
and (\ref{fincollint}). Inspecting these formulae we find that the considered
cross-sections are independent of the electron mass. Consequently there
is no intrinsic mass scale at all and 
$\sigma_{\mathrm{tot}}(\sigma_\Pe=-\sigma'_\Pe)$ is proportional to $E^{-2}$
contributing several per mil to the unpolarized cross-section.

For the evaluation of the hard-photonic bremsstrahlung corrections to
the helicity-conserving channels $(\kappa=\sigma_\Pe=\sigma'_\Pe)$ the 
full phase-space integration including also the non-collinear regions
has to be performed. Instead of applying 
(\ref{masseff}) we prefer to calculate the finite-mass effects using
the results of \refse{collint}. More precisely we exclude the collinear
regions $\theta_i<\Delta\theta_i\ll 1$, $\alpha_i<\Delta\alpha_i\ll 1$ 
($i=1,2$) from the phase-space integration and add the explicit results
of (\ref{incollint}) and (\ref{fincollint}) for the finite-mass effects.
We define the hard-photonic corrections by
\beq
\delta_{\mathrm{HB}} = 
\frac{\left.\si^{\Pem\gamma\to\Pem\gamma\gamma}_{\mathrm{tot}}\right
\vert_{E_{1,2} >\Delta E}}
{\si^{\Pem\gamma\to\Pem\gamma}_{\mathrm{tot}}}
\eeq
and combine them with the virtual and real soft-photonic QED 
corrections $\delta_{\QED}$ of \refse{virc}
\beq
\delta_{\mathrm{QED+HB}} \;=\; 
\delta_{\mathrm{HB}} + \delta_{\QED} \;=\; 
\delta_{\mathrm{HB}} + \delta_{\QED}^{\mathrm{virt}}+\de_{\mathrm{SB}}.
\eeq
Hence the gauge-invariant, IR-finite, and $\Delta E$-independent factor
$\delta_{\mathrm{QED+HB}}$ consists of all photonic corrections to
\egeg. 

$\delta_{\mathrm{QED+HB}}$, which is illustrated in \reffi{eggqm} 
for the single polarization combinations $(\kappa,\lambda_\gamma)$,
turns out to be at the (positive) per-cent level and increases with 
decreasing cuts for $\theta_i$ and $\theta'_\Pe$. Here and in the following
figures the error bars of the Monte Carlo statistics are indicated by 
vertical lines on the corresponding curves. As already observed for
\egezg\ \cite{egnwgezg} the u-channel pole of the cross-sections causes 
a reduction of the numerical accuracy for very small cuts 
$\Delta\theta_\Pe$. However, the case $\Delta\theta_\Pe=1^\circ$ is not 
of physical relevance, but it is considered just for illustrating the 
influence of this cut-off. Since the only dependence on the electron mass 
$\Me$ is logarithmic the photonic corrections are of the form
\beq
\delta_{\mathrm{QED+HB}}=C_1+C_2\log\left(\frac{\Me}{E}\right),
\qquad C_i=\mathrm{const.},\,\, i=1,2,
\eeq
for fixed cut-off parameters $\Delta\theta_\gamma$, $\Delta\theta_\Pe$,
$\delta_\gamma$, $\delta_\Pe$. Consequently the numerical integration 
over the three-particle phase space for non-collinear photon emission 
yields a constant contribution to $\delta_{\mathrm{HB}}$ and has to 
be performed only once for each curve of \reffi{eggqm} as the 
energy-dependent term of $\delta_{\mathrm{HB}}$ is contained in the 
collinearity parts.

\begin{table}
\begin{center}
\arraycolsep 6pt
$$\begin{array}{|c|c||c|c|c||c|c|c|}
\hline
\Delta E/E & \Delta\theta_{1,2}, \Delta\alpha_{1,2} &
\delta_{\mathrm{QED+HB}}/\% & \sigma_{\mathrm{st.dev.}}/\% & \chi^2 &
\delta'_{\mathrm{QED+HB}}/\% & \sigma'_{\mathrm{st.dev.}}/\% &
\chi^{\prime 2} \\
\hline\hline
	& 10^{-3} & 4.05 & 0.11 & 0.23  &   &   &   \\
\cline{2-5}
10^{-3} & 10^{-5} & 4.09 & 0.17 & 0.27  & 4.09 & 0.17 & 0.88  \\
\cline{2-5}
	& 10^{-7} & 4.12 & 0.24 & 0.79  &   &   &   \\
\hline
	& 10^{-3} & 4.01 & 0.17 & 0.37  &   &   &   \\
\cline{2-5}
10^{-5} & 10^{-5} & 4.00 & 0.27 & 0.31  & 4.17 & 0.27 & 0.56  \\
\cline{2-5}
	& 10^{-7} & 4.17 & 0.40 & 0.43  &   &   &   \\
\hline
	& 10^{-3} & 3.94 & 0.24 & 0.58  &   &   &   \\
\cline{2-5}
10^{-7} & 10^{-5} & 4.34 & 0.39 & 0.42  & 4.27 & 0.38  & 0.59  \\
\cline{2-5}
	& 10^{-7} & 4.32 & 0.55 & 0.37  &   &   &   \\
\hline
\end{array}$$
\caption{Cutoff (in-)dependence of the photonic corrections to the
unpolarized cross-section of \protect\egeg\ for
$\protect E_{\mathrm{CMS}}=100\GeV$ and $\Delta\theta_\Pe,
\Delta\theta_\gamma=20^\circ$.}
\label{eggcuttab}
\end{center}
\end{table}
As we have checked analytically as well as numerically all results for
\egezg\ \cite{egnwgezg} hold for \egegg\ after substituting $\MZ\to 0$, 
$g_{\Pe\Pe\PZ}^\kappa\to g_{\Pe\Pe\gamma}^\kappa =1$.
Moreover we have verified that the results for $\delta_{\mathrm{QED+HB}}$
do not depend on the auxiliary IR ($\Delta E$) and angular cuts 
($\Delta\theta_i$, $\Delta\alpha_i$). This fact is demonstrated in 
\refta{eggcuttab}, where the photonic corrections are listed for various
cut-offs together with the corresponding statistical errors and the
$\chi^2$ per degrees of freedom given by {\it VEGAS}. In particular a 
second version for the calculation of these corrections 
($\delta'_{\mathrm{QED+HB}}$) is also included there which follows the 
method of mass-effective factors outlined in \refse{collpho}.

\subsection{Full $\O(\alpha)$ electroweak corrections to \protect\egeg}

\begin{figure}[p]
\begin{center}
\begin{picture}(15,8.5)
\put(-.2,-.5){\includegraphics{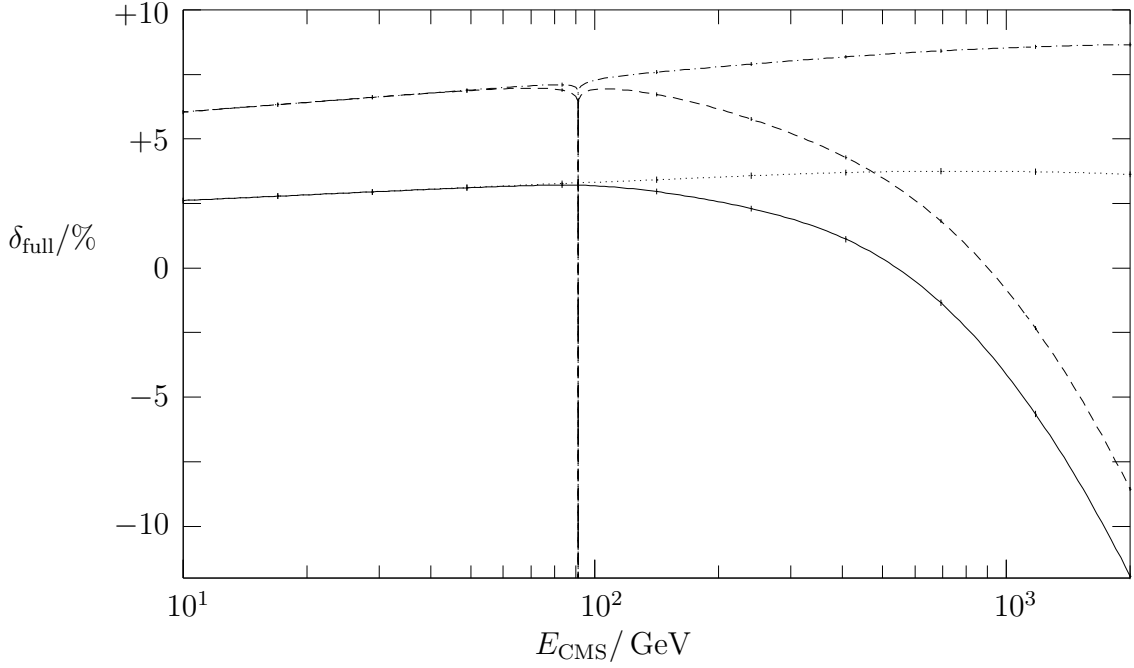}}
\put(0,5.4){$\delta_{\mathrm{full}}/\%$}
\put(7,0){$E_{\mathrm{CMS}}/\GeV$}
\put(2.1,.5){$10^{1}$}
\put(7.6,.5){$10^{2}$}
\put(13.1,.5){$10^{3}$}
\put(1.4,8.4){$+10$}
\put(1.6,6.7){$+5$}
\put(1.6,5.0){$\phantom{-}0$}
\put(1.6,3.3){$-5$}
\put(1.4,1.6){$-10$}
\end{picture}
\end{center}
\caption{Full $\O(\alpha)$ corrections to the integrated cross-section 
($\Delta\theta_\Pe=\Delta\theta_\gamma=20^\circ$) for different polarization 
combinations $(\kappa,\lambda_{\gamma})$ with 
$\kappa=\sigma_\Pe=\sigma'_\Pe$:
----- $(-\frac{1}{2},-1)$, \, $---$ $(-\frac{1}{2},+1)$ \,
$\cdot\cdot\cdot\cdot\cdot$ $(+\frac{1}{2},+1)$, \,
$-\cdot-\cdot-$ $(+\frac{1}{2},-1)$.}
\label{eggax}
\end{figure}

The full $\O(\alpha)$ corrections to Compton scattering are furnished by 
the photonic contributions of the previous section together with the 
virtual electroweak RCs given in \refse{virc}
\beq
\delta_{\mathrm{full}} \;=\;
\delta_{\mathrm{QED+HB}} + \delta_{\mathrm{weak}} \;=\;
\delta_{\mathrm{QED+HB}} + \delta_{\NC} + \delta_{\PW}.
\eeq
Since the results for $\delta_{\mathrm{weak}}$ have already been
discussed in detail in \cite{egeg} we immediately turn to 
$\delta_{\mathrm{full}}$, which is illustrated in \reffi{eggax} for the 
electron-helicity-conserving channels $(\kappa=\sigma_\Pe=\sigma'_\Pe)$.
For energies below the scale of the weak gauge bosons 
$\delta_{\mathrm{full}}^{+\kappa}(+\lambda_\gamma)$ and
$\delta_{\mathrm{full}}^{-\kappa}(-\lambda_\gamma)$ coincide as the
(parity-non-conserving) weak corrections $\delta_{\mathrm{weak}}$ do
not yield sizeable contributions there and we are left with the pure
(parity-conserving) photonic corrections $\delta_{\mathrm{QED+HB}}$.
For energies above the threshold singularity at $E_{\mathrm{CMS}}=\MZ$,
which shows up as a tiny (logarithmic) peak,
considerable (negative) weak corrections arise for left-handed 
electrons which even exceed the (positive) photonic corrections of
pure QED for several hundred GeV. On the other hand 
$\delta_{\mathrm{weak}}^+$ hardly contributes to $\delta_{\mathrm{full}}^+$
for right-handed electrons. Altogether the RCs in $\O(\alpha)$ to the 
polarized Compton cross-sections are of the order $\lsim 10\%$ for 
energies between 10\GeV\ and 2\TeV.
\begin{figure}[p]
\begin{center}
\begin{picture}(15,8.5)
\put(-.2,-.5){\includegraphics{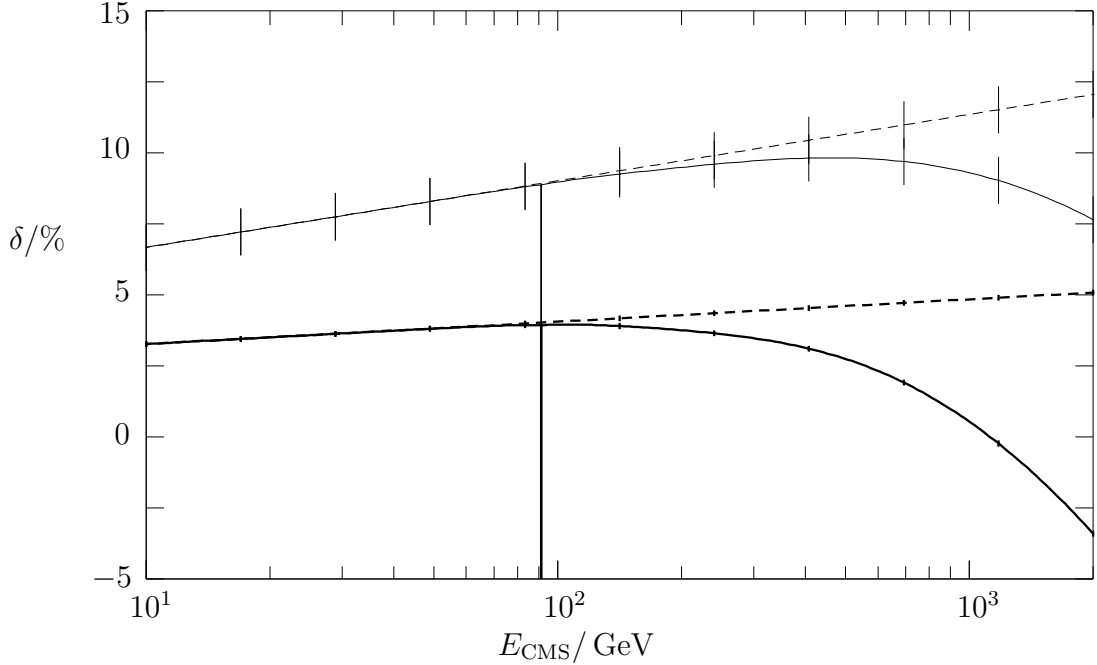}}
\put(0.5,5.4){$\delta/\%$}
\put(7,0){$E_{\mathrm{CMS}}/\GeV$}
\put(2.1,.5){$10^{1}$}
\put(7.6,.5){$10^{2}$}
\put(13.1,.5){$10^{3}$}
\put(1.7,8.4){$15$}
\put(1.7,6.6){$10$}
\put(1.9,4.7){$5$}
\put(1.9,2.8){$0$}
\put(1.6,0.9){$-5$}
\end{picture}
\end{center}
\caption{Full $\O(\alpha)$ and photonic corrections to the unpolarized
cross-section integrated with $\Delta\theta_\gamma=\Delta\theta_\Pe=20^\circ$ 
(thick curves) and $\Delta\theta_\gamma=0^\circ$, $\Delta\theta_\Pe=1^\circ$:
----- $\delta_{\mathrm{full}}$, \, $---$ $\delta_{\mathrm{QED+HB}}$.}
\label{eggau}
\end{figure}

The full $\O(\alpha)$ RCs to the unpolarized cross-section are compared
to the photonic corrections in \reffi{eggau}. While Compton scattering
represents practically a pure electromagnetic process for energies
below $\MZ$ the (negative) weak corrections exceed the (positive) 
photonic corrections in the TeV range. From \reffi{eggau} we can
also deduce that the $\O(\alpha)$ RCs are the more enhanced the more tiny
we choose the angular cuts $\Delta\theta_\gamma$ and $\Delta\theta_\Pe$.
Finally the single contributions of the weak and photonic RCs to
$\delta_{\mathrm{full}}$ are summarized in \refta{eggtab} where we have 
again compared the two versions $\delta_{\mathrm{QED+HB}}^{(\prime)}$ for
the photonic corrections.
\begin{table}
\begin{center}
\arraycolsep 4pt
$$\begin{array}{|c||c|c|c|c|c|c|c|c|}
\hline
E_{\mathrm{CMS}}
& \multicolumn{2}{c|}{50\GeV}
& \multicolumn{2}{c|}{100\GeV}
& \multicolumn{2}{c|}{500\GeV}
& \multicolumn{2}{c|}{2\TeV} \\
\hline
\Delta\theta_\Pe/\Delta\theta_\gamma
& 1^\circ/0^\circ & 20^\circ/20^\circ
& 1^\circ/0^\circ & 20^\circ/20^\circ
& 1^\circ/0^\circ & 20^\circ/20^\circ
& 1^\circ/0^\circ & 20^\circ/20^\circ \\
\hline\hline
\sigma_{\mathrm{Born}}/\pba
& 520.226 & 205.363 & 130.057 & 51.341
& 5.202 & 2.054 & 0.3251 & 0.1284 \\
\hline\hline
\delta_{\mathrm{full}}/\%
& 8.74 & 3.80 & 9.48 & 3.95 & 10.51 & 2.74 & 8.49 & -3.42 \\[-.6em]
& \scriptstyle\pm 0.83 & \scriptstyle\pm 0.11
& \scriptstyle\pm 0.83 & \scriptstyle\pm 0.11
& \scriptstyle\pm 0.83 & \scriptstyle\pm 0.11
& \scriptstyle\pm 0.83 & \scriptstyle\pm 0.11 \\
\hline
\delta_{\mathrm{weak}}/\%
& -0.00 & -0.01 & -0.04 & -0.10 & -0.83 & -1.86 & -4.41 & -8.49 \\
\hline
\delta_{\mathrm{QED+HB}}/\%
& 8.74 & 3.81 & 9.52 & 4.05 & 11.34 & 4.60 & 12.90 & 5.07 \\[-.6em]
& \scriptstyle\pm 0.83 & \scriptstyle\pm 0.11
& \scriptstyle\pm 0.83 & \scriptstyle\pm 0.11
& \scriptstyle\pm 0.83 & \scriptstyle\pm 0.11
& \scriptstyle\pm 0.83 & \scriptstyle\pm 0.11 \\
\hline
\delta'_{\mathrm{QED+HB}}/\%
& 8.42 & 3.80 & 9.6 & 4.09 & 9.8 & 4.62 & 11.2 & 5.14 \\[-.6em]
& \scriptstyle\pm 0.81 & \scriptstyle\pm 0.16
& \scriptstyle\pm 1.1  & \scriptstyle\pm 0.17
& \scriptstyle\pm 1.9  & \scriptstyle\pm 0.18
& \scriptstyle\pm 1.8  & \scriptstyle\pm 0.21 \\
\hline
\end{array}$$
\caption{Full, weak and photonic corrections to \protect\egeg\ for the 
unpolarized cross-section.}
\label{eggtab}
\end{center}
\end{table}

Note that apart from a trivial normalization effect the polarization 
asymmetry of the incoming photon is not influenced by the photonic corrections 
$\delta_{\mathrm{QED+HB}}$ at all. Thus the weakly corrected polarization 
asymmetry, which is discussed in \cite{egeg}, is practically identical with 
the full $\O(\alpha)$ corrected one.

\section{Summary}

Applying the computer-algebra packages {\it FeynArts} \cite{fa} and 
{\it FeynCalc} \cite{fc} we have calculated the virtual electroweak 
radiative corrections to high-energy Compton scattering. In comparison 
with the existing results of \cite{egeg} the obtained analytical 
expressions turn out to be very simple and well-suited for numerical 
evaluations or further theoretical investigations.
Moreover we have given compact results for the helicity amplitudes of
double Compton scattering in terms of Weyl-van der Waerden spinor
products valid for non-collinear photon emission. The inclusion of the
corresponding finite-mass effects of the electron has been described 
in detail.
Finally we have presented numerical results both for the purely photonic
as well as for the full $\O(\alpha)$ corrections to the integrated
cross-section of \egeg\ for polarized and unpolarized particles.
For energies ranging from 10\GeV\ to 2\TeV\ the corrections modify the
lowest-order cross-sections roughly by $5-10\%$.

Together with the investigation of the virtual electroweak radiative
corrections to \egnwezeg\ in \cite{egeg,egnwez} and the radiative 
processes \egnwgezg\ in \cite{egnwgezg} this work completes the
discussion of gauge-boson production in electron-photon collisions
up to the first order in $\alpha$. In view of experimental requirements
further studies on this subject are desirable. In particular the angular 
distributions of the hard-photonic corrected cross-sections as well as 
the energy spectra of the hard photons should also be investigated.

\section*{Acknowledgement}

The author would like to thank A.\ Denner and H.\ Spiesberger for helpful 
discussions and advice.

\appendix
\def\theequation{\thesection.\arabic{equation}}
\setcounter{equation}{0}
\section*{Appendix}

\section{List of scalar integrals}
\label{scalar}

For completeness here we list all scalar one-loop integrals which
have been used for the calculation of the virtual RCs given in
\refse{virc}. Following the conventions of \cite{scalar} the scalar 
integrals are defined by:
\beqar
&& B_{0}(p_{1}^{2},m_{0},m_{1})=
\frac{(2\pi\mu)^{4-D}}{i\pi^{2}}\int d^{D}q
\frac{1}
{[q^{2}-m_{0}^{2}+i\varepsilon]
[(q+p_{1})^{2}-m_{1}^{2}+i\varepsilon]},\nn
\\[1em]
&& C_{0}(p_{1}^{2},(p_{2}-p_{1})^{2},p_{2}^{2},m_{0},m_{1},m_{2})=
\frac{1}{i\pi^{2}}\int d^{4}q \nn \\[.2em]
&& \hspace{1em}\times\frac{1}
{[q^{2}-m_{0}^{2}+i\varepsilon]
[(q+p_{1})^{2}-m_{1}^{2}+i\varepsilon]
[(q+p_{2})^{2}-m_{2}^{2}+i\varepsilon]}, \nn \\[1em]
&& D_{0}(p_{1}^{2},(p_{2}-p_{1})^{2},(p_{3}-p_{2})^{2},p_{3}^{2},
p_{2}^{2},(p_{3}-p_{1})^{2},m_{0},m_{1},m_{2},m_{3})=
\frac{1}{i\pi^{2}}\int d^{4}q \nn \\[.2em]
&& \hspace{1em}\times\frac{1}
{[q^{2}-m_{0}^{2}+i\varepsilon]
[(q+p_{1})^{2}-m_{1}^{2}+i\varepsilon]
[(q+p_{2})^{2}-m_{2}^{2}+i\varepsilon]
[(q+p_{3})^{2}-m_{3}^{2}+i\varepsilon]}.\hspace{3em}
\eeqar
All needed 2-point functions $B_0$, which are calculated in $D$ 
space-time dimensions with $D\to 4$, can be easily derived from the 
special cases:
\beqar
B_{0}(x,m,m) &=& \Delta+2-\log\left(\frac{m^2}{\mu^2}\right)
+\beta_{xm}\log\left(\frac{\beta_{xm}-1}{\beta_{xm}+1}\right),
\qquad \beta_{xm} = \sqrt{1-\frac{4m^2}{x+\ie}},
\nn \\[.2em]
B_{0}(x,0,m) &=& \Delta+2-\log\left(\frac{m^2}{\mu^2}\right)
+\left(\frac{m^2}{x}-1\right)\log\left(1-\frac{x+\ie}{m^2}\right).
\label{b0s}
\eeqar
Of course all renormalized quantities do not depend on the arbitrary 
reference mass $\mu$ and the constant
\beq
\Delta = \frac{2}{4-D}-\gamma_{E}+\log 4\pi
\eeq
which contains the UV divergences.
Except for the photonic contribution to the field-renormalization
constant $\delta Z^{\Pe,\kappa}$ of the electron,
which we have taken from \cite{hab&mex}, the $B_0$ functions of
(\ref{b0s}) are also sufficient for the determination of the
field-renormalization constants which contain 
$\partial B_0/\partial p^2$. In particular the abbreviations 
$B_{t\PW\PW}$, $B_{s0\PW}$, and $B_{u0\PW}$ used in (\ref{wvirc}) 
are given by:
\beqar
B_{t\PW\PW} &=& B_{0}(t,\MW,\MW)-B_{0}(0,\MW,\MW),
\nn \\[.2em]
B_{v0\PW} &=& B_{0}(v,0,\MW)-B_{0}(0,0,\MW),
\qquad v=s,u.
\eeqar
The following 3- and 4-point functions are calculated for the
limit $s,-t,-u,\MW^2\gg\Me^2$ where the infinitesimal photon
mass $\lambda$ regularizes possible IR divergences. Scalar
functions which are related by crossing symmetry 
($s\leftrightarrow u$) are given generically with the abbreviation 
$r=s+\ie,u+\ie$.
\beqar
C_{0}(\Me^2,0,r,0,\Me,\Me) &=& \frac{1}{r}\left[2\zeta(2)
+\frac{1}{2}\log^2\left(\frac{\Me^2}{-r}\right)\right],
\qquad \zeta(2) = \frac{\pi^2}{6},
\nn\\[.2em]
C_{0}(\Me^2,\Me^2,t,\Me,\lambda,\Me) &=& \frac{1}{t}\left[
\frac{1}{2}\log^2\left(\frac{\Me^2}{-t}\right)
+\log\left(\frac{\Me^2}{-t}\right)
\log\left(\frac{\lambda^2}{\Me^2}\right)
-\zeta(2)\right],
\nn\\[.2em]
C_{0}(0,0,t,\Me,\Me,\Me) &=& \frac{1}{2t}
\log^2\left(\frac{\Me^2}{-t}\right),
\nn\\[.2em]
C_{0}(\Me^2,0,r,\MZ,\Me,\Me) &=& \frac{1}{r}\left[
\Li\left(\frac{r}{\MZ^2}\right)
-\log\left(\frac{\Me^2}{\MZ^2-r}\right)
\log\left(1-\frac{r}{\MZ^2}\right)
\right],
\nn\\[.2em]
C_{0}(0,0,t,0,\MZ,0) &=& \frac{1}{t}\left[
\zeta(2)-\Li\left(1+\frac{t}{\MZ^2}\right)
\right],
\nn\\[.2em]
C_{0}(0,0,r,0,\MW,\MW) &=& \rlap{$ C_{r\PW\PW} $} \hspace{3.2em}
= -\frac{1}{r}\Li\left(\frac{r}{\MW^2}\right),
\nn\\[.2em]
C_{0}(0,0,t,\MW,0,\MW) &=& \rlap{$ C_{t\PW\PW} $} \hspace{3.2em}
= \frac{1}{t}\log^2\left(\frac{\beta_{t\PW}+1}{\beta_{t\PW}-1}\right),
\qquad \beta_{t\PW} = \sqrt{1-\frac{4\MW^2}{t}},
\nn\\[.2em]
C_{0}(0,0,t,\MW,\MW,\MW) &=& \rlap{$ C_{t\PW\PW\PW} $} \hspace{3.2em}
= \frac{1}{2t}\log^2\left(\frac{\beta_{t\PW}+1}{\beta_{t\PW}-1}\right),
\eeqar
\beqar
D_0(0,\Me^2,0,\Me^2,r,t,\Me,\Me,\lambda,\Me) &=& \frac{1}{rt}\left[
2\log\left(\frac{\Me^2}{-t}\right)\log\left(\frac{\lambda\Me}{-r}\right)
-3\zeta(2)\right],
\nn\\[.4em]
D_0(0,\Me^2,0,\Me^2,r,t,\Me,\Me,\MZ,\Me) &=& \frac{1}{t(r-\MZ^2)}\left[
-2\log\left(\frac{\Me^2}{-t}\right)\log\left(1-\frac{r}{\MZ^2}\right) 
\right.\nn\\[.2em]
&& \hspace{-8em} \phantom{\times\biggr\{}\left.
+\frac{1}{2}\log^2\left(\frac{-t}{\Me^2}\right)
+\Li\left(1+\frac{t}{\MZ^2}\right)
-4\Li\left(\frac{r}{r-\MZ^2}\right)
-\zeta(2)\right],
\nn\\[.4em]
D_0(0,0,0,0,r,t,\MW,\MW,0,\MW) &=& D_{rt} = 
\frac{1}{\sqrt{t^2(r-\MW^2)^2-4r^2t\MW^2}} 
\;\disp\sum_{n=1}^{2}(-1)^{n+1}
\nn\\[.2em]
&& \hspace{-15em} \times\left[
\log\left(1-\frac{r}{\MW^2}\right)\log(-x_n)
-\Li\left(1+\frac{\beta_{t\PW}+1}{\beta_{t\PW}-1}x_n\right)
-\Li\left(1+\frac{\beta_{t\PW}-1}{\beta_{t\PW}+1}x_n\right)
\right.\nn\\[.2em]
&& \hspace{-15em} \phantom{\times\biggr\{}\left.
+3\Li(1+x_n)
-\Li\left(1+\frac{x_n\MW^2}{\MW^2-r}\right)
-\eta\left(-x_n,\frac{\MW^2}{\MW^2-r}\right)
\log\left(1+\frac{x_n\MW^2}{\MW^2-r}\right)
\right],
\nn\\[.4em]
&& \hspace{-15em} \mbox{with: }\quad
x_{1,2} = \left[t(r-\MW^2)-2r\pm\sqrt{t^2(r-\MW^2)^2-4r^2t\MW^2}\right]
/2(r+t),
\eeqar
The dilogarithm $\Li(x)$ and the $\eta$-function $\eta(x,y)$ are defined
as usual
\beqar
\Li(x) = -\int_0^x\,\frac{dt}{t}\,\log(1-t), 
&&\qquad -\pi<\mathrm{arc}(1-x)<\pi,\\[.2em]
\eta(x,y) = \log(xy)-\log(x)-\log(y),
&&\qquad -\pi<\mathrm{arc}(x),\mathrm{arc}(y)<\pi.
\eeqar

\end{document}